%% file: 0_main.tex
\documentclass[cameraready]{Interspeech}

\title{Align-Consistency: Improving Non-autoregressive and Semi-supervised ASR with Consistency Regularization}

\author[affiliation={}]{Wanting}{Huang}
\author[affiliation={},correspondingauthor]{Weiran}{Wang}

\address{
    Department of Computer Science\\
    University of Iowa\\
    Iowa City, IA, USA
}

\email{\{wanting-huang,weiran-wang\}@uiowa.edu}

\keywords{consistency regularization, non-autoregressive decoding, semi-supervised ASR, self-training}

\def\aa{{\mathbf{a}}}
\def\x{{\mathbf{x}}}
\def\y{{\mathbf{y}}}
\def\tx{\tilde{\mathbf{x}}}

\usepackage{comment}
\usepackage{amsmath,amssymb}
\usepackage{algorithm}
\usepackage{algpseudocode}
\DeclareMathOperator*{\argmax}{arg\,max}

\usepackage{multirow}

\begin{document}

\maketitle

\begin{abstract}
    %
    Consistency regularization (CR) improves the robustness and accuracy of Connectionist Temporal Classification (CTC) by ensuring predictions remain stable across input perturbations. In this work, we propose Align-Consistency, an extension of CR designed for Align-Refine---a non-autoregressive (non-AR) model that performs iterative refinement of frame-level hypotheses. This method leverages the speed of parallel inference while significantly boosting recognition performance. The effectiveness of Align-Consistency is demonstrated in two settings. First, in the fully supervised setting, our results indicate that applying CR to both the base CTC model and the subsequent refinement steps is critical, and the accuracy improvements from non-AR decoding and CR are mutually additive. Second, for semi-supervised ASR, we employ fast non-AR decoding to generate online pseudo-labels on unlabeled data, which are used to further refine the supervised model and lead to substantial gains.
\end{abstract}
    
\input{1_intro}
\input{2_related}
\input{3_method}
\input{4_expts}
\input{5_conclusion}

\bibliographystyle{IEEEtran}
\bibliography{mybib}

\end{document}

%% file: 1_intro.tex
\section{Introduction}
\label{sec:intro}

End-to-end (E2E) automatic speech recognition (ASR) can be roughly divided into categories: autoregressive models such as attention-based encoder-decoder~\cite{chorowski2015attention,chan2016listen} and RNN-Transducer~\cite{graves2012sequence},
and non-autoregressive (non-AR) models, with a prominent example being Connectionist Temporal Classification (CTC,~\cite{graves2006ctc}).
Due to the language modeling component (the decoder) in autoregressive models, decoding is performed in a left-to-right fashion by beam search.
On the other hand, non-autoregressive (non-AR) decoding models aim to update many (or all) positions in parallel, offering a simple and efficient inference procedure~\cite{ghazvininejad2019maskpredict,chi2021alignrefine,wang2022streamingalignrefine}.
Despite strong results, E2E ASR depends heavily on labeled speech, and performance can degrade significantly in low-resource conditions.
This motivates semi-supervised learning that leverages abundant unlabeled speech, 
commonly through self-training with pseudo-labels~\cite{kahn2020selftraining,chen2020e2e_selftraining}.

%

On the other hand, consistency regularization (CR) is a recently proposed method that enforces invariance of model predictions under stochastic perturbations, and it has yielded substantial accuracy improvement for CTC~\cite{yao2025crctc}.
However, its application to other models remains largely under-researched.

Motivated by these developments, we propose to combine benefits of non-AR decoding and strong regularization, and further extend their usage to the semi-supervised setting, for building a stronger ASR model. 
Our contributions are as follows.
\begin{itemize}
    \item 
    We propose Align-Consistency, which extends CR to Align-Refine~\cite{chi2021alignrefine}, a non-AR decoding based on iterative refinements of hypothesis alignment. By enforcing consistency throughout the entire refinement process, we achieve significant accuracy improvement. The same learning objective is applied to the semi-supervised setup, where we initialize from a fully supervised model, and use the fresh model to generate pseudo-labels on the fly, based on which the model is finetuned. The effectiveness of our method is demonstrated through evaluations in two different settings. 
    \item {Supervised:}
    On LibriSpeech, in both low-resource (LS-100) and high-resource (LS-960) settings, iterative refinement and consistency each reduces WER, while combining them yields the best result. Compared with CR-CTC, Align-Consistency improves test WERs from $12.2/26.7$ to $10.0/22.9$ on LS-100, and from $4.3/9.9$ to $3.3/7.4$ on LS-960.

    \item {Semi-supervised:}
    In self-training with unlabeled LibriSpeech 960h or LibriLight 6000h, Align-Consistency greatly benefits from unlabeled data and achieves substantial WER reductions over the initialization, suggesting robustness of CR to noisy supervision. From the LS-100 model, we reduce test WERs from $10.0/22.9$ to $4.3/9.6$ with 960h unlabeled data, and further to $3.8/9.1$ with additional 6000h unlabeled data.
    From the LS-960 model, we reduce test WERs from $3.3/7.4$ to $2.5/5.7$ with 6000h unlabeled data.
\end{itemize}


%% file: 2_related.tex
\section{Related Work}
\label{sec:related}


\subsection{Non-autoregressive decoding and iterative refinement}
\label{sec:related:nar}
\vspace*{-.5ex}

Non-autoregressive (non-AR) decoding avoids strictly left-to-right generation and enables parallel updates over multiple or all output positions.
A common strategy is iterative refinement, where the model revises an initial hypothesis over several steps to gradually incorporate label dependencies and improve accuracy.
Existing non-autoregressive methods can be broadly grouped into those that perform iterative refinement on label sequences versus those focusing on frame-level alignments.
Label-sequence based approach typically follows the mask-predict paradigm~\cite{ghazvininejad2019maskpredict} and iteratively edits low-confidence positions.
Mask-CTC initializes from CTC outputs and refines masked tokens~\cite{higuchi2020maskctc}, and related non-autoregressive transformers perform parallel editing on the label sequence~\cite{chen2021nat}.
Alignment-based approach instead operates on a frame-level token path and supports insertions and deletions without explicit length prediction.
Representative methods include Imputer~\cite{chan2020imputer} and Align-Refine~\cite{chi2021alignrefine}, as well as its streaming extension~\cite{wang2022streamingalignrefine}.

Our method is based on the Align-Refine model, which consists of a CTC module and an additional transformer decoder.
Let $\x$ be an input utterance with $T$ acoustic frames, and let $\mathcal{V}$ denote the token set augmented with the blank. We use $\aa=[a_0,\dots,a_T]$ to denote an alignment sequence of length $T$.
Let $\hat{\aa}^{(0)}$ be the greedy hypothesis from the CTC module, containing the most probably tokens from per-frame posteriors:
\begin{gather}
\label{eq:ar_init}
\hat{{a}}^{(0)}_{t}
=
\argmax_{v\in\mathcal{V}}
\; p_{\mathrm{ctc}}(a_t=v \mid \x),
\qquad t=1,\ldots,T.
\end{gather}
Align-Refine then iteratively predicts an improved frame-level hypothesis using the transformer decoder $\mathrm{Dec}$, conditioned on the previous hypothesis alignment (with full attention context in this work) and the audio features from the encoder shared with CTC: $\mathbf{h}=\mathrm{Enc}(\x)$.
For refinement step $s=1,\dots, S$, we have
\begin{gather}
\label{eq:ar_ps}
p^{(s)} (\aa \mid \hat{\aa}^{(s-1)}, \x)
= \mathrm{Dec} ( \hat{\aa}^{(s-1)}, \mathbf{h} ),\\
\label{eq:ar_update}
\hat{{a}}^{(s)}_{t} = \argmax_{v\in\mathcal{V}} \; p^{(s)} (a_t=v \mid \hat{\aa}^{(s-1)},\x).
\end{gather}
Both the CTC module (considered as step $s=0$, with $\aa^{(-1)}$ being empty) and the refinement steps are trained by maximizing the log-likelihood of label sequence under CTC:
\begin{gather*}
\log p^{(s)} (\y\mid\x)
= \log \sum_{\aa\in \mathcal{B}^{-1} (\y)} p^{(s)} (\aa \mid \hat{\aa}^{(s-1)},\x),
\end{gather*}
where $\mathcal{B}^{-1} (\y)$ is the set of alignments that reduce to $\y$ after reducing repetitions and blanks.
All components of Align-Refine model are trained jointly with the following non-AR loss
\begin{align}
\notag
\mathcal{L}_{\mathrm{nar}} (\x,\y; \theta)
:= &- \alpha \cdot \log p_{\mathrm{ctc}} (\y \mid \x) \\ \label{eq:ar_mle}
& \hspace*{-3em} - (1-\alpha) \cdot \frac{1}{S}\sum_{s=1}^{S}\log p^{(s)} (\y \mid \hat{\aa}^{(s-1)}, \x)
\end{align}
where $\theta$ denotes all parameters of the model.   
Throughout this work, we set $\alpha=0.3$ and $S=2$.

\subsection{Self-training and pseudo-labeling for ASR}
\label{sec:related:selftrain}
\vspace*{-.5ex}

Self-training has a long history in ASR, where unlabeled speech is decoded by a seed model and the resulting pseudo-labels are used as supervision~\cite{wessel2004unsupervised,lamel2002lightly,ma2008unsupervised,yu2010lvcsr}.
With neural acoustic models, selection and agreement criteria were studied to control label noise~\cite{vesely2013semisupervised,vesely2017wordselection,quitry2016agreement}.
For end-to-end ASR, \cite{chen2020e2e_selftraining} proposed an online self-training procedure based on CTC, which repeatedly generates pseudo-labels on mini-batches and updates the model immediately, which can be interpreted as alternating optimization over pseudo-labels and model parameters.
For a dataset with $N_l$ labeled utterances and $N_u$ unlabeled utterances, the  objective of~\cite{chen2020e2e_selftraining} can be written as
\begin{align}
\label{eq:e2e_st_obj}
\min_{\theta,\{\y_j\}_{j=1}^{N_u}}
\frac{1}{N_l}\sum_{i=1}^{N_l}\mathcal{L}(\x_i,\y_i;\theta)
+ \frac{\gamma}{N_u}\sum_{j=1}^{N_u}\mathcal{L}(\x_j,\y_j;\theta),
\end{align}
where $\{\y_j\}$ are pseudo-labels obtained by decoding with the current model (CTC in their case) and held fixed when updating $\theta$.
Related work further showed that strong seed models, stable decoding, and appropriate filtering improve robustness at scale~\cite{khurana2021dust,kahn2020selftraining}.
In this paper, we show that Align-Refine leads to more accurate pseudo-labels than CTC and therefore leads to better ASR performance with self-training.




%% file: 3_method.tex
\section{Method}
\label{sec:method}
\vspace*{-.5ex}

We propose a new learning objective, named Align-Consistency, which integrates alignment-based iterative refinement (Align-Refine) with consistency regularization for end-to-end ASR, and then extend its use to semi-supervised learning.

\subsection{Align-Consistency: Align-Refine with CR}
\label{sec:method:base}
\vspace*{-.5ex}




Let $\tx_1$ and $\tx_2$ be two augmented versions of the clean input, obtained by applying SpecAugment~\cite{park2019specaugment} twice on the same clean input $\x$ (with different random seeds), and let $\y$ be the ground truth label sequence of $\x$. 
We forward the non-AR model on both $\tx_1$ and $\tx_2$, and let $p^{(s)}_1$ and $p^{(s)}_2$ be frame-level posteriors produced by the step $s$ of our non-AR model on $\tx_1$ and $\tx_2$ respectively, for $s=0,\dots,S$, with $s=0$ corresponding to the base CTC model. See~\eqref{eq:ar_ps} for the definition of these posteriors, here we omit alignments and input for brevity.
We use consistency regularization (CR,~\cite{yao2025crctc}) on $p^{(s)}_1$ and $p^{(s)}_2$ to enforce their agreement, with the following symmetric KL loss:
\begin{align}
\notag
\mathcal{L}_{\mathrm{cr}} (p^{(s)}_1, p^{(s)}_2) := &\\ \label{eq:consistency}
& \hspace*{-7em}
\frac{1}{2} \sum_{t=1}^T \left( \mathrm{KL} \left(\mathrm{sg}(p^{(s)}_{1,t}), p^{(s)}_{2,t} \right)
+ \mathrm{KL} \left(\mathrm{sg}(p^{(s)}_{2,t}), p^{(s)}_{1,t} \right) \right)
\end{align}
where $\mathrm{sg}(\cdot)$ denotes the stop-gradient operation. We note that 
the computation of CR does not require the ground truth label and can thus be applied to unlabeled data.

We apply CR to both the
base CTC and the refinement steps in our non-AR model. We refer to the resulting model \emph{Align-Consistency}, with the following loss for supervised learning:
\begin{align} \notag
\mathcal{L}_{\mathrm{AC}} (\tx_1,\tx_2,\y)
:= & \frac{1}{2} \left(\mathcal{L}_{\mathrm{nar}} (\tx_1,\y)
 + \mathcal{L}_{\mathrm{nar}} (\tx_2,\y)
 \right) \\
\label{eq:sup_total}
& \hspace*{-7em} + \lambda_0 \cdot \mathcal{L}_{\mathrm{cr}} (p^{(0)}_1, p^{(0)}_2)
+ \lambda_1 \cdot \frac{1}{S} \sum_{s=1}^S \mathcal{L}_{\mathrm{cr}} (p^{(s)}_1, p^{(s)}_2)
\end{align}
where $\lambda_0$ and $\lambda_1$ denote the CR loss weight for the base CTC and refinement steps respectively.

\subsection{Semi-supervised self-training with Align-Consistency}
\label{sec:method:semi}
\vspace*{-.5ex}

We follow the end-to-end self-training method of~\cite{chen2020e2e_selftraining} for semi-supervised learning.
For each unlabeled utterance $\x$, its pseudo-label $\hat{\y}$ is generated from the \emph{clean} input (to obtain the most accurate decoding result) with our non-AR model. Intuitively, given the refinement steps have lower WER than the base CTC, it is better to use decoding results of the last step as pseudo-labels than that of CTC. This is 
verified by our ablation study (see Sec~\ref{sec:exp:pseudo-label}) and we use the last step hypotheses by default. 

Once $\hat{\y}$ is determined, we compute the Align-Consistency loss on the unlabeled data similarly to~\eqref{eq:sup_total}. That is, we create two augmented versions $\tx_1$ and $\tx_2$, and compute the $\mathcal{L}_{\mathrm{nar}}$ terms using $\hat{\y}$, and the $\mathcal{L}_{\mathrm{cr}}$ terms without labels. 

For a dataset with $N_l$ labeled utterances and $N_u$ unlabeled utterances, our final semi-supervised loss is defined as
\begin{align*}
\min \mathcal{L}_{\mathrm{semi}} := &
\frac{1}{N_l}\sum_{i=1}^{N_l}\mathcal{L}_{\mathrm{AC}} (\tx_{i,1},\tx_{i,2},\y_i) \\
& \hspace*{4em} + \frac{\gamma}{N_u}\sum_{j=1}^{N_u}\mathcal{L}_{\mathrm{AC}} (\tx_{j,1},\tx_{j,2},\hat{\y}_j).
\end{align*}
We find $\gamma=1$ to work well in practice, as observed by~\cite{chen2020e2e_selftraining}. 
Unless otherwise mentioned, we use the same loss weights $(\lambda_0, \lambda_1)$ within both the supervised and unsupervised losses.


%% file: 4_expts.tex
\section{Experiments}
\label{sec:exp}

\subsection{Datasets}
\label{sec:exp:data}
\vspace*{-.5ex}

We evaluate Align-Consistency on LibriSpeech~\cite{panayotov2015librispeech} and Libri-Light~\cite{kahn2020librilight}.
LibriSpeech is a widely used benchmark of approximately 1000 hours of 
read English speech derived from LibriVox audiobooks.
For supervised training, we use two standard labeled subsets: train-clean-100 (LS-100) and train-960 (LS-960).
For semi-supervised self-training, we additionally leverage the 6k-hour subset of Libri-Light (LL-6000) as unlabeled speech.
All models are evaluated on the standard LibriSpeech development and test sets 
using word error rate (WER).

\begin{table}[t]
\centering
\caption{Fully-supervised results on LibriSpeech development sets (WER$\downarrow$). $s=0$ corresponds to the base CTC, $s=2$ uses 2 refinement steps on top of base CTC. Note Align-Consistency reduces to CR-CTC for $\alpha=1.0$.}
\label{tab:fs_ablation}
\begin{tabular}{lccrrrr}
\toprule
\multirow{2}{*}{Dataset} &
\multirow{2}{*}{$\lambda_{\mathrm{0}}$} &
\multirow{2}{*}{$\lambda_{\mathrm{1}}$} &
\multicolumn{2}{c}{$s=0$} & \multicolumn{2}{c}{$s=2$} \\
\cmidrule(lr){4-5}\cmidrule(lr){6-7}
& & & clean & other & clean & other \\
\midrule

\multirow{12}{*}{LS-100} &
\multicolumn{6}{c}{$\alpha=1.0$} \\
& 0.0 & & 12.5 & 27.2 \\
& 0.1 & & 12.2 & 26.6 \\
& 0.2 & & 12.0 & 26.2 \\
& 0.3 & & 12.1 & 26.3 \\
\cmidrule(lr){2-7}
& \multicolumn{6}{c}{$\alpha=0.3$} \\
& 0.0 & 0.0 & 11.6 & 24.9 & 10.4 & 24.0 \\
& 0.2 & 0.0 & 10.8 & 24.4 & 10.0 & 23.2 \\
& 0.0 & 0.2 & 10.7 & 24.3 & 10.2 & 23.0 \\
& \textbf{0.2} & \textbf{0.2} & \textbf{10.3} & \textbf{24.0} & \textbf{9.6} & \textbf{22.5} \\
& 0.1 & 0.3 & 10.8 & 24.5 & 10.1 & 23.4 \\
& 0.3 & 0.1 & 10.9 & 24.2 & 10.3 & 23.0 \\
\midrule

\multirow{12}{*}{LS-960}
& \multicolumn{6}{c}{$\alpha=1.0$} \\
& 0.0 & & 4.5 & 10.4 \\
& 0.1 & & 4.1 & 9.5 \\
& 0.2 & & 3.9 & 9.2 \\
& 0.3 & & 3.9 & 9.5 \\
\cmidrule(lr){2-7}
& \multicolumn{6}{c}{$\alpha=0.3$} \\
& 0.0 & 0.0 & 3.8 & 8.7 & 3.2 & 7.6 \\
& 0.2 & 0.0 & 3.5 & 8.4 & 3.0 & 7.4 \\
& 0.0 & 0.2 & 3.6 & 8.3 & 3.1 & 7.2 \\
& \textbf{0.2} & \textbf{0.2} & \textbf{3.4} & \textbf{8.1} & \textbf{2.9} & \textbf{6.9} \\
& 0.1 & 0.3 & 3.6 & 8.3 & 3.1 & 7.1 \\
& 0.3 & 0.1 & 3.6 & 8.5 & 3.0 & 7.4 \\
\bottomrule
\end{tabular}
\end{table}

\subsection{Modeling Pipeline}
\vspace*{-.5ex}

We follow the ESPnet~\cite{watanabe2018espnet} LibriSpeech recipe for ASR modeling and keep the architecture fixed across all experiments. We use a subword vocabulary of 300 BPEs for LS-100 and 5000 BPEs for LS-960, following the standard recipes.
Our backbone uses a Conformer encoder with 12 layers, attention dimension 512 (8 heads), and a feed-forward dimension of 2048. Our model has about 110M parameters for LS-100 and about 116M parameters for LS-960.
We perform a $2\times$ time reduction at the encoder output before the prediction heads, by average pooling over time with both window size and stride size being 2.
We perform $S=2$ refinement steps over the base CTC module during both model training and inference, since our preliminary study confirms that 2 steps capture most of the WER gain, as also observed by previous work on Align-Refine. 
Unless otherwise noted, the base CTC loss weight $\alpha$ in~\eqref{eq:ar_mle} is set to $0.3$.
For each training scenario, we average checkpoints of the last 5 epochs to obtain the final model parameters for inference.


\subsection{Supervised training on LS-100 and LS-960}
\vspace*{-.5ex}

We train both the LS-100 model and the LS-960 model for 30 epochs.
WER results on development sets are given in Table~\ref{tab:fs_ablation}.

We first reproduce the success of CR for the pure CTC model, which corresponds to our formulation with $\alpha=1$. In this case, we tune $\lambda_0$ over $\{0.1, 0.2, 0.3\}$ and observe 0.2 to work the best, similar to the finding of ~\cite{yao2025crctc}. We obtain relative improvements of $4.0\%/3.7\%$ on LS-100 and $13.3\%/11.5\%$ on LS-960 over pure CTC.

We then switch to the Align-Consistency model which uses $\alpha=0.3$, and tune CR coefficients around 0.2. We provide recognition results of Align-Consistency at two inference steps $s=0$ (corresponding to the base CTC module) and $s=2$.

First note that, even without CR (i.e., $\lambda_0=\lambda_1=0$), base CTC ($s=0$) outperforms the $\alpha=1$ configuration: on LS-100, dev WER improves from $12.5/27.2$ to $11.6/24.9$; on LS-960, it improves from $4.5/10.4$ to $3.8/8.7$, indicating that the refinement steps help with learning a better encoder than CTC alone. Integrating additional refinement steps further enhances accuracy by more effectively capturing label dependencies.

Regarding CR coefficients, enabling CR on either CTC alone (e.g., $(\lambda_0, \lambda_1)=(0.2,0.0)$) or on refinement steps alone (e.g., $(\lambda_0, \lambda_1)=(0.0,0.2)$) already leads to sizable WER gains.
And using CR to both of them with $(\lambda_0,\lambda_1)=(0.2,0.2)$ achieves the best performance.
On LS-100, these optimal CR coefficients reduce WERs from 10.4/24.0 to 9.6/22.5 on dev-clean/dev-other, corresponding to a relative improvement of 7.7\% and 6.2\%, respectively. On LS-960, the optimal CR coefficients reduce WERs from 3.2/7.6 to 2.9/6.9, yielding 9.4\%/9.2\% relative reductions. 

In summary, our supervised experiments demonstrate that Align-Consistency outperforms CTC as a non-AR method. Furthermore, CR reinforces all internal components of the model, yielding significant complementary gains in accuracy.

\begin{table}[t]
\centering
\caption{Self-training results (WER$\downarrow$) on LS-100+LS-960, evaluated on dev sets. We test two pseudo-label (PL) sources: base CTC vs. Align-Consistency. 
We initialize from the best model trained on LS-100.}
\label{tab:st_100_960_pl_source}
\begin{tabular}{lccccc}
\toprule
\multirow{2}{*}{$(\lambda_{\mathrm{0}},\lambda_{\mathrm{1}})$} &
\multirow{2}{*}{\parbox{3em}{\# Inf.\\steps $s$}} &
\multicolumn{2}{c}{CTC PL} &
\multicolumn{2}{c}{Align-Cons. PL} \\
\cmidrule(lr){3-4}\cmidrule(lr){5-6}
& & clean & other & clean & other \\
\midrule
Initialization  & 0 & 10.3 & 24.0 & {10.3} & {24.0} \\
(no PL) & 2 & 9.6 & 22.5 & {9.6} & {22.5} \\
\midrule
\multirow{2}{*}{(0.0, 0.0)}
& 0 & 7.1 & 15.5 & 6.8 & 15.2 \\
& 2 & 6.6 & 14.4 & 6.3 & 14.2 \\
\midrule
\multirow{2}{*}{(0.2, 0.0)}
& 0 & 6.8 & 14.9 & 6.5 & 14.4 \\
& 2 & 6.2 & 13.8 & 6.0 & 13.5 \\
\midrule
\multirow{2}{*}{(0.0, 0.2)}
& 0 & 6.7 & 15.1 & 6.4 & 14.4 \\
& 2 & 6.2 & 13.7 & 5.8 & 13.2 \\
\midrule
\multirow{2}{*}{(0.2, 0.2)}
& 0 & 6.4 & 14.2 & 6.1 & 14.0 \\
& 2 & 5.9 & 13.3 & \textbf{5.5} & \textbf{12.6} \\
\bottomrule
\end{tabular}
\end{table}

\subsection{Semi-supervised self-training}
\vspace*{-.5ex}

Starting from a supervised checkpoint, we introduce unlabeled speech and continue with self-training as described in Sec~\ref{sec:method:semi}. We run self-training for 110 epochs with the LS-100 model, and 80 epochs for the LS-960 model.

\subsubsection{Choice of pseudo-label}
\label{sec:exp:pseudo-label}
\vspace*{-.5ex}

We first test the two choices for pseudo-label generation, using either the base CTC or the final step of Align-Consistency. 
This ablation study is performed on the setup where LS-100 is used as supervised set while LS-960 is used as unsupervised.
For either choice, we tune the CR coefficients and give WERs at decoding steps $s=0$ and $s=2$. The WER results are shown in Table~\ref{tab:st_100_960_pl_source}.
As expected, pseudo-labels generated by Align-Consistency are more useful than those generated by base CTC, across all CR coefficients.
Using $(\lambda_0,\lambda_1)=(0.2,0.2)$ for both supervised and unsupervised loss achieves the best performance, and we reduce the WERs of Align-Consistency from 9.6/22.5 to 5.5/12.6, a $42.7\%/44.0\%$ relative improvement.

\subsubsection{CR on unsupervised data}
\label{sec:exp:cr-unsup}
\vspace*{-.5ex}

One might wonder, since CR was initially proposed for fully-supervised training, whether it will be useful on unsupervised data. We perform an ablation study where we remove the CR on unsupervised data during self-training. This choice leads to 
significant WER degradations, as shown in Table~\ref{tab:self_training_selected}. These results show that CR is hugely beneficial even when used in combination with noisy pseudo-labels.

\begin{table}[t]
\centering
\caption{Ablation study on the use of CR loss on unsupervised data for self-training on LS-100+LS-960, evaluated on dev sets. Pseudo-labels are generated by Align-Consistency.}
\label{tab:self_training_selected}
\begin{tabular}{ccrrrr}
\toprule
Sup CR & Unsup CR &
\multicolumn{2}{c}{$s=0$} & \multicolumn{2}{c}{$s=2$} \\
\cmidrule(lr){3-4}\cmidrule(lr){5-6}
$(\lambda_0, \lambda_1)$ & $(\lambda_0^\prime, \lambda_1^\prime)$ & clean & other & clean & other \\
\midrule
\multirow{2}{*}{(0.2, 0.2)} & (0.2, 0.2) & 6.1 & 14.0 & 5.5 & 12.6 \\
& (0.0, 0.0) & 6.7 & 14.8 & 6.2 & 13.9 \\

\bottomrule
\end{tabular}
\end{table}

\begin{table}[t]
\centering
\caption{Self-training results on LibriSpeech dev sets (WER$\downarrow$), with different amounts of unsupervised data and initializations.}
\label{tab:self_training}
\begingroup
\begin{tabular}{llrrrr}
\toprule
\multirow{2}{*}{Data} & \multirow{2}{*}{$(\lambda_0, \lambda_1)$} &
\multicolumn{2}{c}{$s=0$} & \multicolumn{2}{c}{$s=2$} \\
\cmidrule(lr){3-4}\cmidrule(lr){5-6}
& & clean & other & clean & other \\
\midrule
LS-100 
& (0.2, 0.2) & 10.3 & 24.0 & 9.6 & 22.5 \\
\midrule
\multirow{2}{*}{+LS-960}
& (0.0, 0.0) & 5.6 & 11.6 & 5.0 & 10.9 \\
& (0.2, 0.2) & 4.6 & 10.7 & 4.1 & 9.5 \\
\midrule
\multirow{2}{*}{\parbox{5em}{+(LS-960 \\+LL-6000)}}
& (0.0, 0.0) & 4.9 & 11.2 & 4.4 & 10.6 \\
& (0.2, 0.2) & \textbf{4.1} & \textbf{10.3} & \textbf{3.6} & \textbf{8.8} \\
\midrule
\midrule
LS-960
& (0.2, 0.2) & 3.4 & 8.1 & 2.9 & 6.9 \\
\midrule
\multirow{2}{*}{+LL-6000}
& (0.0, 0.0) & 3.0 & 6.6 & 2.7 & 5.7 \\
& (0.2, 0.2) & \textbf{2.6} & \textbf{6.1} & \textbf{2.2} & \textbf{5.2} \\
\bottomrule
\end{tabular}
\endgroup
\end{table}

\begin{table}[t]
\centering
\caption{WERs results (\%) on LibriSpeech test sets.
}
\label{tab:self_training_test}
\begin{tabular}{llrr}
\toprule
\multirow{2}{*}{Model} & \multirow{2}{*}{Parameters} &
test- & test-  \\
& & clean & other \\
\midrule
\multicolumn{4}{c}{LS-100 Sup} \\
LPM~\cite{hsu2020lpm} &  & 14.9 & 40.0 \\
\hspace*{1em} + LS-860h &     & 8.4 & 22.9 \\
\hspace*{2em}  + LL-60000h &   & 7.2 & 20.8 \\
\multicolumn{2}{l}{IPL~\cite{xu2020ipl} + LS-860h}    & 6.0 & 10.3 \\
\multicolumn{2}{l}{\hspace*{2em} + LibriVox-54Kh}  & 5.1 & 8.8  \\
NST~\cite{park2020nst} &  & 5.5 & 16.9 \\
\hspace*{1em} + LS-860h &     & 4.5 & 9.5 \\
\midrule
CTC & ($\alpha\!=\!1$, $\lambda_0\!=\!0.0$) & 12.8 & 27.6 \\
CR-CTC & ($\alpha\!=\!1$, $\lambda_0\!=\!0.2$) & 12.2 & 26.7 \\
Align-Consistency & ($\lambda_0\!=\!0.2$, $\lambda_1\!=\!0.2$) & 10.0 & 22.9 \\
\hspace*{0.6em} + LS-960 &  &4.3 & 9.6 \\
\multicolumn{2}{l}{\hspace*{2.1em} + (LS-960 + LL-6000)} & 3.8 & 9.1 \\
\midrule
\midrule
\multicolumn{4}{c}{LS-960 Sup} \\
\multicolumn{2}{l}{IPL~\cite{xu2020ipl} + LibriVox-54Kh} & 2.2 & 4.7 \\
NST~\cite{park2020nst} &  & 1.9 & 4.1 \\
\hspace*{1em} + LL-60Kh &  & 1.7 & 3.7 \\

\midrule
CTC & ($\alpha\!=\!1$, $\lambda_0\!=\!0.0$) & 4.7 & 10.8 \\
CR-CTC & ($\alpha\!=\!1$, $\lambda_0\!=\!0.2$)  & 4.3 & 9.9 \\
Align-Consistency & ($\lambda_0\!=\!0.2$, $\lambda_1\!=\!0.2$)  & 3.3 & 7.4 \\
\hspace*{1.3em} + LL-6000 & & 2.5 & 5.7 \\
\bottomrule
\end{tabular}
\end{table}

\subsubsection{More unsupervised data and better initialization}
\label{sec:exp:more-data-better-init}
\vspace*{-.5ex}

We then increase the amount of unsupervised data by adding LL-6000, and also initialize self-training with the 
model trained on LS-960.
The experimental results are given in Table~\ref{tab:self_training}.

As expected, more unsupervised data leads to better accuracy after self-training, with or without CR. In all scenarios, CR provides additional WER gains, and $(\lambda_0,\lambda_1)=(0.2,0.2)$ remains the strongest configuration in each setup.
With CR, the additional 7000h of unsupervised speech improves the dev WERs from 9.6/22.5 to 3.6/8.8 over the LS-100 model; using 6000h of unsupervised speech improves dev WERs from 2.9/6.9 to 2.2/5.2 over the LS-960 model.

\subsection{Final results}
\vspace*{-.5ex}

Finally, Table~\ref{tab:self_training_test} summarizes test WERs of models trained with different amounts of data. Across all settings, Align-Consistency consistently outperforms CR-CTC, justifying our motivation to extend CR to stronger non-AR models. Additionally, Align-Consistency boosts ASR accuracy by effectively using unlabeled speech while maintaining high efficiency in on-the-fly pseudo-labeling. We also compare with representative semi-supervised ASR baselines from the literature under LM-free evaluation. LPM~\cite{hsu2020lpm} and NST~\cite{park2020nst} are based on attention-based encoder-decoder architecture, while IPL~\cite{xu2020ipl} is based on a CTC-trained acoustic model with language model fusion for pseudo-label generation. Despite using only 6000 hours of unlabeled data, we achieve competitive LM-free WERs compared to prior systems trained with one magnitude larger unlabeled data and a much simpler non-AR model, highlighting strong data efficiency.



%% file: 5_conclusion.tex
\section{Conclusion}
\vspace*{-.5ex}

We have presented Align-Consistency, which combines non-AR decoding with consistency regularization for end-to-end ASR.
Our method encourages stable predictions with respect to input perturbations in every inference step, on both clean and noisy labels.
Our experiments on both fully supervised and semi-supervised settings show that non-AR decoding and consistency regularization are complementary. 
In the future, we plan to extend the same principles to broader class of non-AR models, including recently developed diffusion LLM-based ASR models~\cite{wang2025audioconditioned}.